\documentclass[aps,prd,twocolumn,nofootinbib]{revtex4}

\usepackage{latexsym}
\usepackage{amsmath}
\usepackage{amssymb}
\usepackage{eufrak}
\usepackage{euscript}
\usepackage{epsfig,graphics,graphicx}

\newcommand{\bel}[1]{\begin{equation}\label{#1}}
\newcommand{\bal}[1]{\begin{eqnarray}\label{#1}}

\newcommand{\be}{\begin{equation}}
\newcommand{\ee}{\end{equation}}
\newcommand{\ba}{\begin{eqnarray}}
\newcommand{\ea}{\end{eqnarray}}

\newcommand{\bes}{\begin{equation*}}
\newcommand{\ees}{\end{equation*}}

\newcommand{\txi}{\tilde{\xi}}
\newcommand{\trc}{\tilde{R}_c}

\begin{document}

\title{Phase conversion in a weakly first-order quark-hadron transition}

\author{A. {\sc Bessa}$^{1}$\footnote{abessa@if.usp.br},
E. S. {\sc Fraga}$^{2}$\footnote{fraga@if.ufrj.br} 
and B. W. {\sc Mintz}$^{2}$\footnote{mintz@if.ufrj.br}}

\affiliation{$^{1}$Instituto de F\'\i sica, Universidade de S\~ao Paulo,
Caixa Postal 66318, 05315-970, S\~ao Paulo, SP , Brazil \\
$^{2}$Instituto de F\'\i sica, Universidade Federal do Rio de Janeiro, 
Caixa Postal 68528, 21941-972 Rio de Janeiro, RJ , Brazil}

\begin{abstract}
We investigate the process of phase conversion in a thermally-driven {\it weakly} 
first-order quark-hadron transition. This scenario is physically appealing even if 
the nature of this transition in equilibrium proves to be a smooth crossover for 
vanishing baryonic chemical potential. We construct an effective potential by 
combining the equation of state obtained within Lattice QCD for the partonic sector 
with that of a gas of resonances in the hadronic phase, and present numerical results 
on bubble profiles, nucleation rates and time evolution, including the effects from 
reheating on the dynamics for different expansion scenarios. Our findings confirm the 
standard picture of a cosmological first-order transition, in which the process of phase 
conversion is entirely dominated by nucleation, also in the case of a weakly first-order 
transition. On the other hand, we show that, even for expansion rates much lower than 
those expected in high-energy heavy ion collisions, nucleation is very unlikely, 
indicating that the main mechanism of phase conversion is spinodal decomposition. 
Our results are compared to those obtained for a strongly first-order transition, as the 
one provided by the MIT bag model.
\end{abstract}


\maketitle

\section{Introduction}

It is widely accepted that experiments in high-energy heavy ion collisions at the 
Relativistic Heavy Ion Collider (RHIC) have produced clear signals that nuclear matter 
undergoes a phase transition to a deconfined partonic phase at sufficiently large values 
of energy density \cite{rhic,Muller:2006ee}. A similar transition presumably took place in 
the early universe a few seconds after the big bang \cite{cosmo,Schwarz:2003du}. In fact, 
this whole picture is expected from quantum chromodynamics (QCD), which exhibits the 
phenomenon of asymptotic freedom. The nature of the quark-hadron transition, nevertheless, 
remains an open question. Although Lattice QCD \cite{Laermann:2003cv}, whose recently 
improved techniques allow for performing calculations with almost realistic quark masses 
\cite{Aoki:2006br}, seems to indicate a crossover, the possibility of a weakly first-order 
transition is not ruled out from the experimental point of view. In reality, most hydrodynamic 
calculations within high-energy heavy ion collisions adopt an equation of state which provides 
a strongly first-order transition \cite{rhic}.

Another point that is seldom mentioned is that, being built on equilibrium assumptions, 
Lattice QCD thermodynamics does not provide any information on the dynamical nature of the 
deconfining transition. In actual experiments, the critical (dynamical) behavior could be 
very different from what one would expect from a crossover in the (equilibrium) phase diagram, 
simply because the phase conversion is achieved via a nonequilibrium evolution 
process \cite{dynamics,reviews}. Indeed, results from simulations in statistical mechanics show 
that the critical behavior can differ significantly if one compares the equilibrium phase diagram 
to the (nonequilibrium) time evolution of a given system, and suggest that the dynamics could  
change the nature of the phase transition, even though the situation is still unclear \cite{statistical,berg}. 
For these reasons, the scenario of a {\it weakly} first-order deconfining transition is physically appealing, 
especially since it comes out naturally by matching two equations of state that are realistic in their 
own regimes of temperature, namely the equations of state provided by Lattice QCD and by a 
hadron resonance gas.

In this paper we investigate the process of phase conversion in a thermally-driven {\it weakly} 
first-order quark-hadron transition. We build an effective potential by combining the equation of 
state obtained from lattice simulations for one heavy and two light flavors of quarks, which we use 
for the partonic sector, with the equation of state of a gas of resonances for the hadronic phase. 
Using standard techniques for the dynamics of first-order transitions, we compute bubble nucleation 
features, such as bubble profiles, critical radii, the surface tension and the free 
energy as functions of the temperature. We study the process of phase conversion 
evaluating the nucleation rate and investigating the time evolution of the temperature of 
the system and its hadronic fraction, as well as the role played by reheating.

Bubble nucleation is one of different simplified mechanisms used to describe the dynamics 
of a first-order phase transition \cite{reviews}. In this kind of transition, for temperatures 
slightly lower than the critical temperature, $T_{c}$, the thermodynamic potential exhibits a 
metastable minimum besides the global minimum. The former gradually disappears as the system 
cools down, and is turned into a point of inflection known as spinodal instability. By 
decreasing further the temperature, the transition follows a qualitatively different (explosive) 
evolution, since the free energy barrier disappears and there is no extra surface cost 
for small amplitude long-wavelength fluctuations. Therefore, depending on the dynamics of 
supercooling, the phase conversion can proceed via more than one mechanism. In a slowly 
expanding system, the phase transition occurs through the nucleation of bubbles of the 
``true vacuum'' state via thermal activation. In general, in order to know whether the 
system reaches the spinodal instability before nucleation is completed, it is necessary to 
investigate the time scale for thermal nucleation relative to that for the expansion. In 
the case of the quark-gluon plasma (QGP) presumably formed in ultrarelativistic 
heavy ion collisions, the expansion of the plasma is faster than that in the case of the 
early universe during the cosmological quark-hadron transition by a factor of $\sim 10^{18}$. 
Thus, it is clear that one should expect major differences in the process of phase conversion 
when comparing the big bang to the little bang \cite{Yagi:2005yb}.

Our findings confirm the standard picture of a cosmological first-order transition, in which 
the process of phase conversion is entirely dominated by nucleation, also for the case of a 
weakly first-order transition. On the other hand, we show that, even for expansion rates much 
lower than those expected in high-energy heavy ion collisions, nucleation is very unlikely, 
indicating that the main mechanism of phase conversion is spinodal decomposition. Our results 
are compared to those obtained for a strongly first-order transition, as the one provided by 
the MIT bag model.

The dynamics of nucleation and spinodal decomposition in the hadronization of an expanding 
QGP after a high-energy heavy ion collision has been studied under different approaches for 
more than twenty years \cite{Reinhardt:1985nq,Csernai:1992tj,Csernai:1992as,Kapusta:1994pg,
Csorgo:1994dd,Zabrodin:1998dk,Scavenius:1999zc,Shukla:2000dx,Scavenius:2001bb,
polyakov,Fraga:2004hp}, as is also the case for the 
cosmological quark-hadron transition \cite{Enqvist:1991xw,Megevand:2007sv}.
However, previous studies  were mostly based on equations of state obtained in the frame of the 
MIT bag model or within effective models, such as the linear sigma model \cite{GellMann:1960np}, 
the Nambu--Jona-Lasinio model \cite{Klevansky:1992qe} and the Polyakov loop 
model \cite{Pisarski:2000eq} in the environment of a strongly first-order transition. Only 
more recently, stimulated by the findings of Lattice QCD, a few studies have considered 
the case of a smooth crossover, and how its dynamics compares to a strongly first-order 
transition, including the effects from fluctuations and inhomogeneities and the presence of 
a critical point \cite{paech,Aguiar:2003pp,Sasaki:2007db}. Our proposal in this paper is, as 
described above, to investigate a scenario which is in-between these two extrema. 
Furthermore, in the case of high-energy heavy ion collisions, in which there is a very fast 
quench down to the spinodal instability, the dynamics can possibly be well described 
by spinodal decomposition even in the case of a smooth (but very fast) crossover.

The paper is organized as follows. Section II reviews the theoretical methods we use to describe 
the nucleation process of an expanding system, including a brief presentation of the thin-wall 
approximation following a different reasoning, and discusses the dynamics in an expanding 
background plus the role of reheating. Our numerical results are discussed in Section III, 
where we display the relevant bubble features, as well as the time evolution of the temperature 
and of the hadronic fraction, and analyze the process of reheating. Finally, our conclusions are 
presented in Section IV.

\section{Theoretical framework}

Thermodynamically, a phase transition happens when a given system shifts its state of 
equilibrium from one free energy minimum to another in response to the change of some 
critical thermodynamic parameter. In real transitions, this shift between equilibrium 
states is often an essentially  non-equilibrium process and, in principle, one has no 
hope to describe it using the machinery of equilibrium thermodynamics. However, several 
natural systems exhibit phase transitions in which the system is trapped in a 
metastable state (false vacuum) for a long time before reaching the final equilibrium 
configuration (true vacuum). This is certainly true in a strongly first-order scenario 
in which the time scale of supercooling (superheating) in a thermally-driven transition 
is not much faster than the typical reaction time of the system. A first-order phase 
transition occurs between non-contiguous states in the thermodynamic configuration space, 
and manifests itself as a discontinuity in the entropy. This latent heat is a consequence of 
an energy barrier that prevents the system to simply roll down to the true minimum. If this 
barrier is high (characteristic of a strongly first-order transition), a statistically 
improbable fluctuation is required, and the system is held in a metastable state for an 
appreciable time interval.

A similar metastable behaviour sets in when the change in the critical parameter is fast as 
compared to the relaxation time of the system. This is the case of the hadronization of the 
QGP formed in high-energy heavy ion collisions, where the time scale of the expanding plasma 
is orders of magnitude shorter than the typical reaction time of the deconfined matter. It 
turns out that it can be impossible (and immaterial) to determine the ``real nature'' of the 
phase transition in this dynamical system since the quenched plasma does not have to follow 
a continuous path in the thermodynamic configuration space. Assuming a weakly first-order 
scenario for the quark-hadron transition, we review in the next sections some theoretical 
tools used in the description of the decay of metastable states. 

\subsection{Homogeneous nucleation}

The paramount example of metastability in a first-order transition is the supersaturated 
vapor in a gas-liquid mixture below the condensation temperature. The mixture is composed 
of bubbles of liquid suspended in the gas, the two phases separated by a definite surface. 
Taking into account the surface energy of the partially developed phase, the notion of 
thermodynamic equilibrium can be extended in spite of the fact that the gas can not be in 
equilibrium with a fully developed liquid phase. Since it takes energy to make a surface, 
small bubbles will not be energetically favorable, and will be constantly created and 
evaporated. On the other hand, if a sufficiently large bubble is formed in the medium, 
it will tend to grow, eventually converting the entire system to the new phase.

The mechanism of bubble nucleation can play an important role in a first-order phase 
transition. It ignores the initial stages of the development of bubbles, being an 
effective theory for semimacroscopic elements of volume. Still, one may have a flavor 
of what really happens by studying the dynamics of the nucleated bubbles. The question 
of the formation of bubbles is extremely relevant, mainly when external agents play the 
role of nucleating centers, leading to a significant increase of the nucleation rate. 
This is actually what happens to most natural systems, and is known as heterogeneous 
nucleation. However, we will not consider this kind of nucleation mechanism, but one in 
which bubbles originate from intrinsic thermodynamic fluctuations: the mechanism of 
homogeneous nucleation. It is a more fundamental process for which there is a field 
theory which captures its basic features. Most of the formalism is based on a series of 
papers written by Langer in the late sixties \cite{Langer:1967ax}, where the basic 
theoretical apparatus to describe the decay of a metastable state of a classical system 
interacting with a heat bath at temperature $T$ is proposed (see also \cite{langer-turski}). 
In this approach, stable and metastable phases appear as local minima of a smooth energy 
functional $E$. Based on a phenomenological droplet model, it is conjectured that, in 
going from one minimum to a neighbor one, the system is likely to pass across a saddle 
point which is a minimum of $E$ in all directions of functional space but one, the latter 
giving rise to the instability. This saddle point configuration plays the role of the 
critical bubble in the formalism. Langer derived an equation of motion for the probability 
distribution of the system, and obtained a steady-state solution flowing across the 
saddle point, allowing for the calculation of a classical nucleation rate (probability per 
unit time per unit volume that a critical bubble nucleates)
\begin{equation}\label{decayrate1}
 \Gamma = \frac{|\lambda_{-1}|}{\pi T}\; \hbox{Im}\,  F \;,
\end{equation}
where $F$ is a prescribed analytic continuation of the free energy of the stable phase in 
the unbroken (single-well) version of the theory which becomes metastable after the analytic 
continuation. The parameter $\lambda_{-1}$ is the negative eigenvalue of the fluctuation operator 
characterizing the instability of the saddle point configuration.

The zero-temperature quantum field theoretic  case, where transitions are exclusively due to 
quantum tunneling, was considered by Coleman and Callan in \cite{Coleman:1977py}, and, when 
both quantum and thermal fluctuations act together, the decay rate is approximately given by 
(\ref{decayrate1}), as proved by Affleck \cite{Affleck:1980ac} in a quantum mechanical context. 
Of course, the existence of a single direction of instability simplifies the extension of 
Langer's formalism to quantum field theories, and direct applications were proposed in the 
literature \cite{Linde:1980tt,Gleiser:1993hf}.

A good description of the nucleation process relies on a suitable choice of the free energy 
functional governing the dynamics of the bubbles. The natural choice corresponds to deriving 
the free energy from a more fundamental theory. When this is not feasible, and this is the 
case for the full QCD lagrangian, one can resort to a phenomenological approach, imposing 
symmetry requirements, as will be discussed in the next section.

\subsection{Effective potential and equations of state}\label{effVandEoS}

As customary, one can obtain information about the phase transition by studying the evolution 
of a scalar field $\phi$ which represents the order parameter. It is reasonable to assume 
spherical symmetry for nucleating bubbles, so that one defines a coarse-grained free energy 
functional of the form:
\begin{equation}\label{freeenergy}
 F(\phi) = 4\pi \int r^2 dr\,\left[\frac{1}{2}\left(\frac{d\phi}{dr}\right)^2 + V(\phi) \right ] \;. 
\end{equation}
Thus, the field $\phi$ evolves in space in the presence of an effective potential that can 
be parametrized in the form of a Landau expansion around the equilibrium phases, i.e.
\begin{equation}\label{effpot}
 V(\phi) = a(T)\phi^2 - bT\phi^3 + c\phi^4 \;.
\end{equation}
The parameters $b>0$, $c>0$ and $a(T)$, and the interpretation of the order parameter are 
determined by the scenario under consideration, namely the supercooled ($T<T_c$) QGP. 
One should notice that, in this simple approach, the temperature enters only as a parameter 
of the effective potential. 

The potential (\ref{effpot}) is suitable for a first-order phase transition due to the 
properties of its extrema. The order parameter configurations (bubble profiles) are solutions 
of the following Euler-Lagrange equation:
\begin{equation}\label{eulerlagrange}
 \frac{d^2\phi}{dr^2} + \frac{2}{r}\frac{d\phi}{dr} - V'(\phi) = 0 \;.
\end{equation}
The potential (\ref{effpot}) has two minima, $\phi_q$ and $\phi_h$, which correspond to the 
equilibrium phases. Here, $\phi_q$ ($\phi_h$) corresponds to the quark (hadron) phase. There 
is a barrier separating $\phi_q$ and $\phi_h$ which can be associated with a latent heat, a 
jump in entropy from one phase to the other. In addition, one can prove that, as required by 
Langer's formalism, this theory has a saddle-point solution, $\phi_b$, connecting the two minima 
and with a single unstable direction \cite{Langer:1967ax}. The minimum $\phi_q$ is conveniently 
chosen to be zero, while the other one is located at
\begin{equation}
 \phi_h = \frac{1}{8c} \left (3bT + \sqrt{9b^2T^2-32a(T)c}\right) \;.
\end{equation}
The transition temperature is reached when the pressure of the competing phases coincide. 
This condition sets a connection with thermodynamics through the identification 
\begin{equation}\label{thermoconnection}
 p_q(T)-p_h(T) = V(\phi_h,T)\;\;\; ; \;\;\;V(\phi_q,T) =0 \;,
\end{equation}
where $p_q$ and $p_h$ are the pressures of the quark (deconfined) phase and hadron phase, 
respectively. 

In order to proceed with the study of the phase conversion from the deconfined state 
to hadrons, one has to fix the potential, either by connecting it to pressures 
computed for each phase, as described above, or by extracting the effective potential 
directly from some effective field theory, such as the linear sigma model as, for 
instance, in Ref. \cite{Scavenius:2001bb, paech}. We choose to follow the first procedure and, 
instead of using the MIT bag model as usually done for simplicity, we profit from the 
currently more robust knowledge of the equation of state of QCD in the two different 
regimes (partonic and hadronic) which justifies the use of more realistic 
expressions for the pressure. Concretely, we use Lattice QCD results for $N_f =2+1$ quark 
flavors to describe the high-temperature sector \cite{Karsch:2003vd}, 
and a gas of over 250 free resonances for the hadronic phase \cite{KodamaEoS}. This 
yields a {\it weakly} first-order deconfining transition, to be contrasted to the usual case 
of a strongly first-order transition as provided by the bag model, and the value of the 
critical temperature is automatically determined by the crossing 
of the high and low temperature pieces of the equation of state (EoS), as illustrated 
in Fig. 1. For comparison, we also use the bag model for the quark phase in 
our calculations and discussion, choosing the bag constant according to the critical 
temperature obtained by the crossing of the pressure curves \cite{Bessa:2005cm}.

\begin{figure}[!hbt]
\begin{center}
\vskip -0.25 cm
\includegraphics[width=6cm,angle=270]{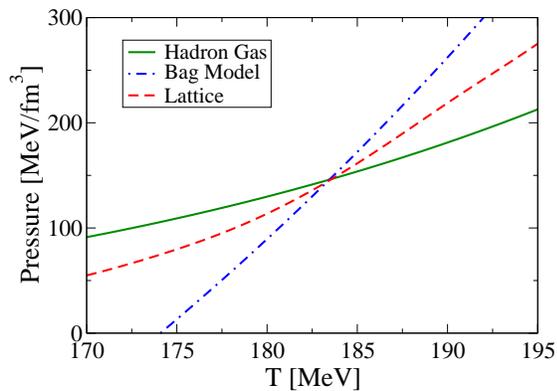}
\caption{\small Pressures for the hadron resonance gas (solid line), bag model (dashed-dotted 
line), and $N_f=2+1$ Lattice QCD (dashed line) equations of state.}
\end{center}
\label{fig:pressure}
\end{figure}

Finally, the relations (\ref{thermoconnection}) leave two remaining conditions to be imposed 
on the potential. These can be chosen to fit a pair of physical properties of the critical 
bubble near the transition temperature, such as the surface tension and the correlation 
length  \cite{Csernai:1992tj}. 

\subsection{The thin-wall approximation revisited}\label{thinwall}

It is always convenient to consider the following auxiliary mechanical problem in the 
solution of Eq. (\ref{eulerlagrange}):  the one-dimensional dynamics of a particle moving 
in a dissipative medium in the presence of the potential $(-V)$. Ignoring the dissipative 
term, strictly at $T=T_c$, the solution of Eq. (\ref{eulerlagrange}) can be written in terms 
of elementary functions. The aim of this section is to discuss approximate analytic solutions 
for temperatures close (in some precise sense) to $T_c$, where dissipation must be present. 

In order to explore this case, it is convenient to write the potential $V$ as the sum of the 
critical potential plus a linear term which introduces deviations from $T=T_c$ 
\cite{Scavenius:2001bb}. This is always possible for a quartic potential up to a shift in 
the zero of the energy. In the present case, we obtain $V(\phi) = W(\phi + \mu) + W_0$, where 
$W_0$ is a constant, 
\begin{equation}
W(\phi) \,=\, c\,(\phi^2 - \phi_0^2)^2 + j\phi\;, 
\end{equation}
\begin{gather}
 \phi_0^2 = \frac{1}{2}\left[\frac{3}{8}\left(\frac{bT}{c}\right)^2 -\frac{a(T)}{c}\right ]\;,\\
 j = \frac{c}{2}\,\left[\frac{a(T)bT}{c^2}-\frac{1}{4}\left(\frac{bT}{c}\right)^3 \right ]\;, 
\end{gather}
and $\mu = -bT/4c$. The corresponding Euler-Lagrange equation has the same form as 
(\ref{eulerlagrange}), with $W'$ replacing $V'$. The critical ($T=T_c$) potential 
(for which $j=0$) is a symmetric double well with minima at $\phi =\pm \phi_0$. In 
terms of these parameters, the solution in the thin-wall approximation at $T=T_c$ is 
a kink which interpolates between the two symmetric minima:
\begin{equation}
 \phi_b(r) \,=\,\phi_0\tanh\left( \frac{r-R}{\xi}\right) \;,
\end{equation}
where $\xi = 1/(\phi_0\sqrt{2c})$ can be thought of as a correlation length. When 
dissipation is neglected, the previous function is a solution for any value of $R$. 
Given the specific form of the dissipation term, approximate solutions for the full 
potential are obtained for large values of $R/\xi$. The limit $\xi/R \ll 1$ characterizes 
the so-called thin-wall regime \cite{reviews}. 

For $T < T_c$, the parameter $j$ is negative and the two minima of $W$ are shifted to
\begin{equation}
 \phi_{f} = \mu \approx -\phi_0 \left(1 +\frac{j}{8c \phi_0^3}\right) 
\end{equation} 
and
\begin{equation}
 \phi_{t} = -\frac{1}{2}\mu + \frac{1}{2}\sqrt{4\phi_0^2 -3\mu^2} 
 \;\approx\; \phi_0 \left(1 -\frac{j}{8c \phi_0^3}\right) \;.
\end{equation} 
When  $|j/(8c\phi_0^3)| \ll 1$, the potential is only slightly tilted. In this range, the dissipation 
term will be small, and it is reasonable to assume that the interpolating solution still has a kink-like 
profile, i.e.:
\begin{equation}\label{thinwallsolution}
 \phi_b(r) \,=\,\phi_f + \frac{\phi_t-\phi_f}{2}\,\left[1-\tanh \left(\frac{r-R}{\xi}\right)\right] \;,
\end{equation}
where $\xi =\sqrt{2/c}(\phi_t-\phi_f)^{-1}$. Still in this limit, one can show that the free energy of the 
bubble $\phi_b$ is given by
\begin{equation}
 F(\phi_b) = -\frac{4\pi R^3}{3}\Delta W + 4\pi R^2 \sigma \; ,
\end{equation}
where $\Delta W = W(\phi_f)-W(\phi_t) \approx 2a|j|$, and the surface tension, $\sigma$, is 
related to the parameters of the problem through
\begin{equation}
 \sigma \equiv \int_{0}^{\infty}\left (\frac{d\phi_b}{dr}\right)^2\,dr \;\approx \; \frac{2}{3\,c\,\xi^3}\;.  
\end{equation}
As discussed in the previous subsection, the value of $\sigma$ at $T=T_{c}$ and the constant value 
of $\xi$ will be considered as inputs in our treatment. In terms of these quantities, the coefficients of the 
potential (\ref{effpot}) are completely determined. Imposing that $\phi_b$ be stationary, we obtain 
the critical radius
\begin{equation}
 R_c = \frac{2\sigma}{\Delta W}\;.
\end{equation}

Strictly speaking, the function (\ref{thinwallsolution}) is an exact solution only at $T=T_c$, where
 $j=0$ and $R_c$ diverges. A different approach consists in taking the dissipation into account in 
 the following manner: (i) the solution conserves energy; (ii) it is exact at $T_c$. We can impose both 
 conditions to a function of the form
\begin{equation}
 \psi_b(r) \,=\,\phi_f + \gamma\,\left[ 1-\tanh \left(\frac{r-\trc}{\txi}\right)\right] \;.
\end{equation}
Condition (ii) gives 
\begin{equation}
\trc = \frac{2\gamma/\txi}{-j+ 4c\, (\mu+\gamma)\left[\phi_0^2-(\mu+\gamma)^2\right ]}\;,
\end{equation}
whereas condition (i) is equivalent to:
\begin{equation}
 \int_0^\infty\frac{2}{r}\left(\frac{d\psi_b(r)}{dr}\right)^2\,dr \;=\;W(\phi_f) -W(\psi_b(0))\;.
\end{equation}
In the thin-wall limit $(\txi/\trc \ll 1)$, condition (i) is simply
\begin{align}
 \frac{8\gamma^2}{3\txi \trc} \;=\; W(\phi_f) -W(\phi_f +2\gamma)\;.
\end{align}
Coupling (i) with (ii) leads to $\gamma = (\phi_v-\phi_f)/2$, $\txi = \xi$ and $\trc=R_c$. 
Numerical results point to a kink-like solution with a 
decreasing escape value when one decreases the temperature. This fact explains the behavior 
of the surface tension plotted in Fig. 4 within this approximation. 

Once we have an (approximated) expression for the bubble solution, we can use it to calculate
 the decay rate shown in Eq. (\ref{decayrate1}). This rate can be written as \cite{Langer:1967ax}
\begin{equation}\label{decayrate2}
 \Gamma = \frac{{\cal P}_0}{(2\pi)}\,e^{-\Delta F/T}\;,
\end{equation}
where $\Delta F$ is the difference in free energy between the saddle-point configuration 
($\phi_b$) and the metastable phase $(\phi_f)$. The prefactor ${\cal P}_0$ is a product of a 
dynamical factor  $\kappa$ (related to the expansion rate of the bubbles) and a statistical factor 
$\Omega_0$, which accounts for the first corrections to $\Delta F$ due to quadratic 
fluctuations around each extremum. As usual, these quadratic fluctuations are formally 
written in terms of the determinant of the fluctuation operator 
$[\nabla^2 - V''(\phi)]$, where $\phi$ is either $\phi_f$ or $\phi_b$. Thus, one obtains:
\begin{equation}\label{omega0}
 \Omega_0 = \left (\pi R_c^2\,T\right)^{1/2}\,\left (\frac{4\pi R_c^2 \,\sigma}{3}\right )^{3/2}\,\left \{\frac{\det^\prime [\nabla^2 - V''(\phi_b)]}{\det [\nabla^2 - V''(\phi_f)]}\right\}^{-1/2} \,,
\end{equation}
where the prime indicates that the zero and negative modes are excluded. The first term 
in the r.h.s. of (\ref{omega0}) comes from the negative eigenvalue of the fluctuation 
operator along the direction of instability. The second term is the contribution from the 
three zero modes which are present because the bubble breaks translational symmetry. Except 
for the four unpaired eigenvalues of $[\nabla^2 - V''(\phi_f)]$, all the other delocalized 
eigenvalues of that operator cancel the corresponding ones of $[\nabla^2 - V''(\phi_b)]$. 
Taking all the remaining contributions into account, it is possible to show that \cite{Linde:1980tt}
\begin{equation}\label{omega_0}
 \Omega_0 = \frac{2}{3\sqrt{3}}\left(\frac{\sigma}{T}\right)^{3/2}\,
 \left(\frac{R_c}{\xi_q}\right)^{4}\;,
\end{equation}
where $\xi_q$ is identified as the correlation length of the metastable phase. 
For the dynamical coefficient, Csernai and Kapusta derived the following 
expression \cite{Csernai:1992tj}:
\begin{equation}\label{kappa}
 \kappa = \frac{4\sigma (\zeta + 4\eta/3)}{(\Delta \omega)^2 R_c^3}\;.
\end{equation}
In the previous formula, $\zeta$ and $\eta$ are respectively the bulk and shear viscosity coefficients 
of the quark phase, and $\Delta \omega$ is difference in the enthalpy density of quark and hadron 
phases (see also Ref. \cite{Venugopalan:1993vk} for a careful discussion of the dynamical 
prefactor). 

\subsection{Dynamics of the phase conversion in an expanding background}

As soon as a QGP is formed after a high-energy collision of heavy ions or in the 
early universe, it expands towards the empty space around it. Due to this expansion, 
the energy density and, therefore, the temperature of the plasma drops and eventually 
becomes smaller than the critical temperature for the quark-hadron phase transition. 
Assuming a weakly first-order transition, as discussed before, the system then becomes 
metastable and the nucleation of bubbles of the cold phase is possible. If the 
expansion rate is large enough, however, the system supercools so fast that it may reach a 
thermodynamically unstable region of the phase diagram, the spinodal region, before 
nucleation is able to drive most of the system to the true vacuum. In this case, the phase 
conversion is dominated by the process of spinodal decomposition, which is rather 
different from nucleation, since it is characterized by long-ranged fluctuations,
instead of localized nucleating bubbles \cite{reviews}. Here we 
focus on the nucleation stage during the expansion, and how the nucleation and 
expansion time scales compare.

For simplicity, we consider a homogeneous and isotropic expansion of the fluid, and 
assume that the Hubble parameter $H(x) \equiv \partial_\mu u^\mu(x)$ is actually a constant. 
We also assume that the entropy is approximately conserved (globally) during the expansion, 
once the viscosity of the QGP is very low, as indicated by heavy-ion collision experiments 
\cite{Muller:2006ee, Muller:2007rs}. However, the viscosity can not be strictly zero, once 
it is necessary for the process of nucleation at vanishing chemical potential (see, e.g., 
ref. \cite{Venugopalan:1993vk}). For definiteness, we assume that the viscosity coefficient 
in eq. (\ref{kappa}), $\zeta + 4\eta/3$, is of the order of the lower bound set by holographic 
models \cite{Kovtun:2003wp}. 

To have a measure of the importance of nucleation in the hadronization process, 
we estimate how much of an expanding plasma should hadronize through nucleation before the 
spinodal temperature is reached. In a first approach, we employ some simplifying assumptions. 
The first of them is to neglect reheating effects (which will be considered later) 
and, assuming the conservation of entropy: 
\begin{equation}
 \frac{d}{dt}(sa^3) = 0 \;.
\end{equation}
The assumption of isotropic expansion leads to the Hubble law for the scale factor $a(t)$, 
\begin{equation}\label{eq:Hubble}
 \frac{da}{dt}(t) = Ha(t) \;,
\end{equation}
and Eq. (\ref{eq:Hubble}) leads to an exponential decay of the temperature with time
\begin{equation}
 T(t) = T_c \exp(-Ht) \;,
\end{equation}
where $t=0$ is chosen to correspond to $T=T_c$.

Given an equation of state and an effective potential, we can find the spinodal temperature 
of the system, $T_{sp}$, and the time $t_{sp}$ the plasma takes to reach this temperature. For 
our equation of state, $T_{sp}/T_c = 0.984$, which leads to $Ht_{sp} \sim 0.01$. The fraction 
$f(t)$ of space which suffers nucleation from $t(T_c)\equiv 0$ until $t_{sp} \equiv t(T_{sp})$ 
can be {\it overestimated} as follows: 
\begin{eqnarray}
 f(t_{sp}) &=& \int_0^{t_{sp}}dt\,\frac{4\pi}{3}R^3(t,t_{sp})\Gamma[T(t)] \cr\cr
           &<& \frac{4\pi}{3}(t_{sp}T_c)^4 \sim 10^{-3} \ll 1 \;,
\end{eqnarray}
where $R(t',t)$ is the radius at time $t$ of a bubble ``born'' at time $t'$ [$R(t',t)<c(t-t')$]. 
In this estimate we also use the value $H^{-1} = 10~$fm/c, which should also give an 
extra overestimating contribution to $f(t_{sp})$ 
\footnote{$H^{-1}$ is usually estimated to be in the 
range $1-10~$fm/c for the expanding QGP created in heavy-ion collisions 
\cite{Csorgo:2002ry}.}. 
We see that, if we neglect reheating effects, typical QGP expansion times in heavy ion collisions 
rule out nucleation and therefore most of the plasma should hadronize via spinodal decomposition, 
a result that was also obtained using the linear sigma model \cite{Csorgo:1994dd,Scavenius:2001bb}. 

However, once a bubble of the true vacuum is nucleated, an amount of latent heat proportional 
to the volume of the bubble is released in the medium, so that the temperature does not fall 
exponentially as in the previous case. If the nucleation rate is high enough, the released latent 
heat may win the competition against the energy loss due to the fluid expansion and the plasma 
reheats. This reheating can drive the system to temperatures close to $T_c$, decreasing the 
supercooling rate and considerably delaying its arrival at the spinodal temperature. In this 
case, when we may say that reheating is effective, the whole system is hadronized via nucleation 
of bubbles and $T_{sp}$ will be reached only some time after the transition is completed.

\subsection{Reheating}

In order to account for reheating effects, we make some assumptions aside from those 
cited on the previous sections. First, the latent heat released 
in the formation of a true vacuum bubble is uniformly distributed throughout the whole 
plasma, which is consistent with that of bubble growth via weak deflagration 
\cite{Enqvist:1991xw}. Therefore, if $s_q(T)$ ($s_h(T)$) is the entropy in the QGP (hadron 
gas) phase and $s(T)$ is the space average of the entropy density, 
\begin{equation}
 s = s_hf + (1-f)s_q \;,
\end{equation}
then entropy conservation implies
\begin{equation}
 s = \left(\frac{a(0)}{a(t)}\right)^3 s_q(T_c) \;,
\end{equation}
where we set $T=T_c$ and $s=s_q$ at $t=0$. For a supercooling $\delta\equiv(T_c-T)/(T_c-T_{sp}) \ll 1$, 
we have $s_q(T)\approx s_q(T_c)$ and using the thermodynamic relation $-p = \Delta V = e - Ts$, we 
find a relation between the temperature and the scale factor at a given time $t$:
\begin{equation}\label{eq:T(t)}
 \left(\frac{T(t)}{T_c}\right)^3 = \left(\frac{a(0)}{a(t)}\right)^3 + f \frac{\Delta s(T)}{s_q(T_c)} \;,
\end{equation}
where $\Delta s(T)=s_q(T)-s_h(T)$ is proportional to the latent heat $\ell$ at $T\approx T_c$. 
Once we consider a constant Hubble parameter $H$, the scale factor grows exponentially, 
$a(t) = a(0)e^{Ht}$, and Eq. (\ref{eq:T(t)}) becomes an implicit equation for the temperature 
as a function of time. We see that the first term on the r.h.s. of Eq. (\ref{eq:T(t)}) accounts 
for the cooling due to the expansion, while the second is proportional to the hadronized fraction 
$f$ and reflects the homogeneous reheating caused by the phase conversion.

We compute $f$ using the expression \cite{Guth:1981uk}
\begin{equation}\label{eq:frac_t}
 f(t) = 1 - \exp\left\{-\int_0^t dt' \left(\frac{a(t')}{a(t)}\right)^3 
 \Gamma[T(t')] \frac{4\pi}{3}R^3(t',t)\right\} \;,
\end{equation}
where $R(t',t)$ is the radius of a bubble created at time $t'$ with critical size $R_c[T(t')]$ 
at time $t$:
\begin{equation}\label{eq:R_t}
 R(t',t) = R_c[T(t')]\frac{a(t)}{a(t')} + \int_{t'}^t dt''\,v_w[T(t'')]\frac{a(t)}{a(t'')} \;.
\end{equation}
We suppose that the velocity of the bubble wall at the temperature $T$ is given by 
the following Allen-Cahn equation, which relates the velocity of a domain wall to 
the local curvature \cite{reviews,Fraga:2003mu}:
\begin{equation}\label{eq:v_w}
 v_w(T) = \frac{\Delta p(T)}{\eta} = -\frac{\Delta V(T)}{\eta} \;,
\end{equation}
where $\Delta p(T) = -\Delta V(T) \equiv V_q(T)-V_h(T)>0$ is the pressure difference 
between the two phases at a given temperature $T$. This expression corresponds to a steady 
growth of the bubble in which the pressure difference between both sides of the bubble wall 
is balanced by a damping force which is proportional to the velocity of the wall. The friction 
coefficient $\eta$ is given by 
\begin{equation}
 \eta = {\tilde\eta}T\sigma(T) \;,
\end{equation}
where
\begin{equation}
 \sigma(T) = \int_0^\infty dr\,\left(\frac{d\phi}{dr}\right)^2
\end{equation}
is the surface tension as a function of the temperature, and $\tilde\eta$ is a number of 
order one. 

We solve the set of equations (\ref{eq:T(t)})--(\ref{eq:v_w}) numerically, and the results are 
presented in the following section.

\section{Results and discussion}

\subsection{Bubble features}

In what follows, we calculate the main physical attributes of critical bubbles 
both numerically and using the thin-wall approximation. Fig. 2 shows bubble
profiles, $\Phi(r)$, for various temperatures. It is clear that the numerical results 
and the thin-wall approximation give very similar results for temperatures close 
to $T_c$ and become more and more different as the temperature is lowered, 
as expected. One of the main differences is the broadening of the numerical profile, 
which does not happen with the thin-wall result, indicating not only the failure of this 
approximation for lower temperatures but also the failure of the nucleation picture 
itself as the system approaches the spinodal temperature.

\begin{figure}[!hbt]
\label{fig:profiles}
\begin{center}
\vskip -0.25 cm
\includegraphics[width=6cm,angle=270]{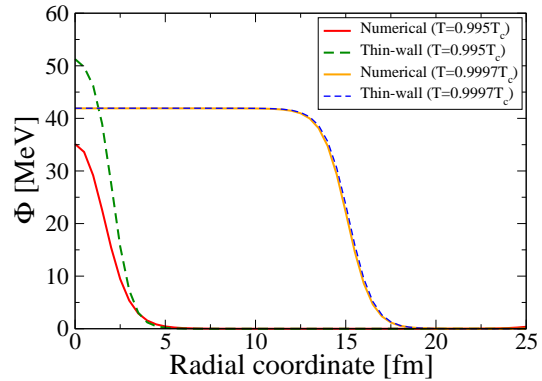}
\caption{Order parameter distributions (bubble profiles), where the 
broken (hadronic) phase is inside the bubble ($r<R_c$) and the unbroken (QGP) 
outside ($r>R_c$).}
\end{center}
\end{figure}

From the bubble profiles we obtain two important quantities: the bubble critical radius and 
the surface tension. The critical radius as a function of the temperature is shown in 
Fig. 3. Notice that the critical radius (defined here as the distance from the bubble center 
in which $\phi=\phi_0/2$) diverges at the critical temperature, 
as seen in Section \ref{thinwall}, and also diverges at the spinodal temperature. Notice, however, 
that for these high values of supercooling, the very concept of a bubble does not make sense any 
longer, once the width of the wall is larger than the bubble radius.

\begin{figure}[!hbt]
\begin{center}
\vskip -0.25 cm
\includegraphics[width=6cm,angle=270]{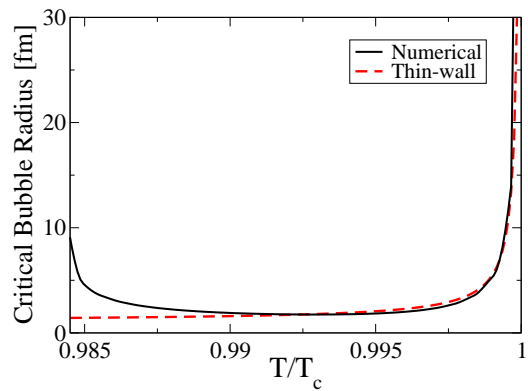}
\caption{Radius of the critical bubble as a function of $T/T_c$.}
\end{center}
\label{fig:critradius}
\end{figure}

The surface tension, $\sigma(T)$, exhibits an interesting behavior, as displayed in Fig. 4. 
From its value at $T=T_c$, $\sigma(T_c)=2\,{\rm MeV/fm}^2$, downwards it starts 
to grow up to a maximum value and then decreases from this point down. Roughly speaking, 
the value of the surface tension is the slope $p(r)\equiv(d\phi/dr)^2$ times the width of the wall 
$\xi(T)$, i.e., the region in which $p(r)\not=0$. As the temperature is lowered from $T_c$, the 
wall is progressively broadened, but $p(r)$ does not change much. As a 
consequence, the surface tension increases. It is only when $p(r)$ starts to decrease 
faster than the increase of $\xi(T)$ that $\sigma(T)$ decreases too. This happens for 
lower temperatures because the order parameter deep into the bubble, $\phi(r=0)$, becomes 
smaller, as can be seen in Fig. 2 and, therefore, $p(r)$ also becomes 
smaller. Notice that the thin-wall approximation is not very sensitive to this competition 
between the width of the wall and the discontinuity of the order parameter across it 
because its bubble profiles always connect the two minima, having $\phi_h(T)$ as the decisive 
parameter, which decreases more slowly than the exact (numeric) $\phi(r=0)$ for lower temperatures. 
Therefore, in the thin-wall approximation, $\sigma(T)$ is a monotonically decreasing function of $T$. 
This nontrivial behavior of $\sigma(T)$ suggests that the temperature in which 
$\sigma(T)$ reaches its maximum can be interpreted as a (generous) limit to the 
applicability of the thin-wall approximation. Following the previous criterion, one can 
say that the thin-wall approximation fails for supercooling higher than $\delta=0.1$.

\begin{figure}[!hbt]
\label{fig:sigma}
\begin{center}
\vskip -0.25 cm
\includegraphics[width=6cm,angle=270]{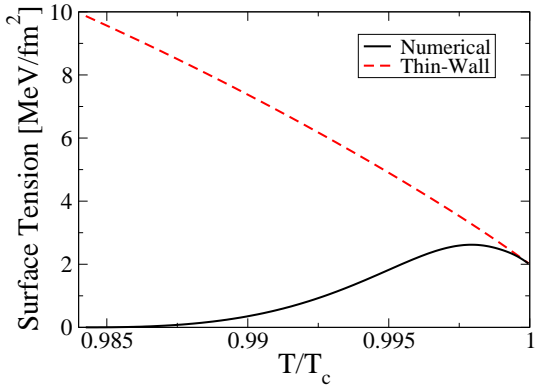}
\caption{Surface tension as a function of $T/T_c$.}
\end{center}
\end{figure}

Our aim in calculating all these quantities is to evaluate the change in free energy 
$\Delta F$ due to the presence of the bubble, which is an essential ingredient for the 
nucleation rate $\Gamma$ (see Eq. (\ref{decayrate2})). $\Delta F$ can be calculated either 
directly from (\ref{freeenergy}) or, in a computationally faster way, by using the formula
\begin{equation}\label{eq:pseudothin-S3}
\Delta F(T) = \frac{4\pi}{3}R_c^3(T)\Delta p(T) + 4\pi\sigma(T)R_c^2(T) \;.
\end{equation}

This expression resembles the thin-wall expression for $\Delta F$, but here the temperature 
dependence of the surface tension makes this formula a good approximation to the exact value 
obtained using  (\ref{freeenergy}), once the functions $R_c(T)$, $\Delta p(T)$ and $\sigma(T)$ are 
known. The different results to $\Delta F$, calculated using a simple thin-wall (where 
$\sigma$ is a constant), using Eq. (\ref{eq:pseudothin-S3}) (which we call {\it T-dependent thin-wall}), 
and from Eq. (\ref{freeenergy}), are shown in Fig. 5. In our following computations, we use Eq. 
(\ref{eq:pseudothin-S3}).

\begin{figure}[!hbt]
\label{fig:action}
\begin{center}
\vskip -0.25 cm
\includegraphics[width=6cm,angle=270]{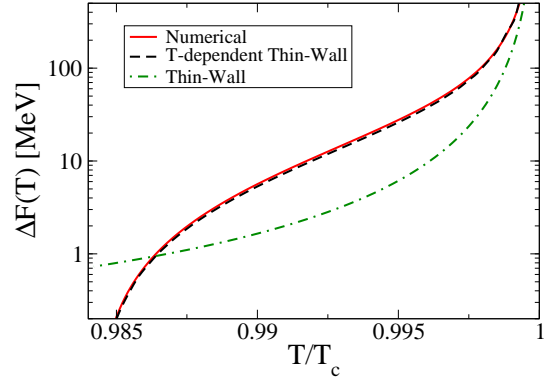}
\caption{Change in free energy, $\Delta F$, calculated in three different ways 
(see text for details). The thin-wall approximation leads to an {\it overestimate} 
of the nucleation rate.}
\end{center}
\end{figure}

\subsection{Time evolution}

Now that we have all the necessary elements, we may investigate the time evolution of the 
system. We first analyze how the temperature evolves with time if we consider the reheating 
backreaction effects on the expanding system. As a first example, we consider three different 
expansion rates: $H^{-1} = 100$ fm/c (fast), $H^{-1} = 600$ fm/c (critical) and $H^{-1} = 10^7$ 
fm/c (slow). From Fig. 6 we can see that in the first case the temperature drops so fast that 
the system reaches the spinodal temperature without having time to grow enough bubbles to 
reheat. This is an example of a {\it quench} into the spinodal region. As the expansion rate $H$ 
is lowered, the growing bubbles eventually have time to effectively release enough latent heat. 
Then, the cooling process is reversed and the temperature is raised to a value close to 
(and lower than) $T_c$. In this way the phase transition is completed through bubble nucleation. 
When the whole plasma is hadronized, there is no more release of latent heat, the expansion once 
again dominates and the temperature falls abruptly. For the realistic equation of state we 
adopt, we find that the lowest (critical) expansion rate $H$ at which there is reheating corresponds  approximately to $H^{-1}=600$ fm/c. Finally, we also consider a very low expansion and verify that 
the phase conversion proceeds almost entirely at $T=T_c$, which is equivalent to having an almost 
{\it in equilibrium} phase conversion. Notice, however, that, if we take a closer look at the 
temperature as a function of time, even for the slowest expansion rate there is a slight 
supercooling in the beginning of the phase transition (as can be seen in the zoom for early 
times displayed in Fig. 7). In any case, the slower the expansion rate, the closer to 
equilibrium the transition evolves. Furthermore, the product $Ht_{sp}$ is the same for all 
expansion rates. 

\begin{figure}[!hbt]
\label{fig:reaq1ads}
\begin{center}
\vskip -0.25 cm
\includegraphics[width=6cm,angle=270]{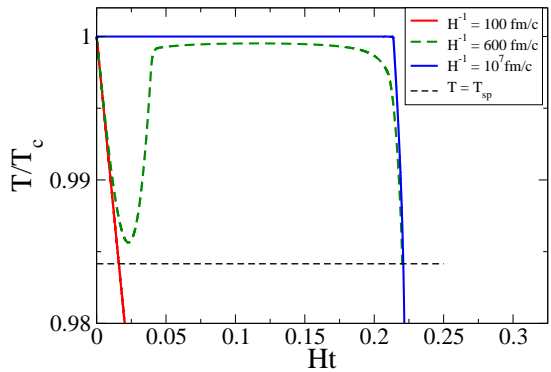}
\caption{Temperature, in units of $T_{c}$, as a function of time, in units of $1/H$, 
for different expansion rates.}
\end{center}
\end{figure}

Dissipation effects can generally affect significantly the dynamics of phase conversion, 
even in the case of an explosive spinodal decomposition \cite{Fraga:2004hp}. 
In our framework, viscosity enters as a parameter in the dynamical prefactor of the 
nucleation rate (see Eq. (\ref{kappa})). Recalling that the viscosity of the QGP is presumably very 
small \cite{Muller:2006ee, Muller:2007rs} and that, on the other hand, holographic models 
set a lower bound ($\eta_{AdS}/s=1/4\pi$) \cite{Kovtun:2003wp}, we compare the dynamics 
of the phase conversion for two values of viscosity: $\eta_{AdS}$ and $3\eta_{AdS}$. Fig. 8 shows that
by increasing the viscosity from $\eta_{AdS}$ to $3\eta_{AdS}$ the supercooling decreases approximately
 by a factor of two, although the time to complete the transition remains unaffected.

\begin{figure}[!hbt]
\label{fig:reaq1ads_zoom}
\begin{center}
\vskip -0.25 cm
\includegraphics[width=6cm,angle=270]{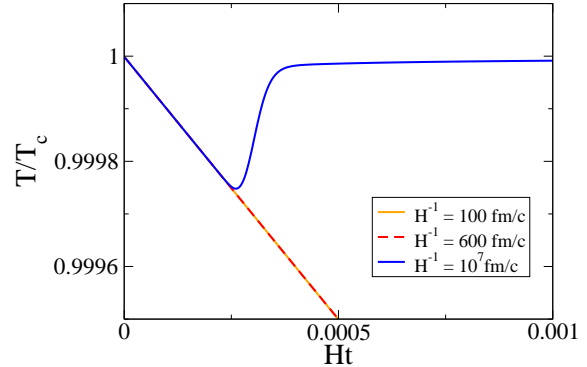}
\caption{Temperature, in units of $T_{c}$, as a function of time, in units of $1/H$, 
for different expansion rates at early times.}
\end{center}
\end{figure}

The effects of different values of latent heat on the dynamics of the phase transition 
was also studied. In Fig. 9, we show the time evolution of 
the temperature considering the two equations of state discussed in 
Section \ref{effVandEoS} and an expansion rate $H^{-1}=600$ fm/c. 

\begin{figure}[!hbt]
\label{fig:reaq600_1_3ads}
\begin{center}
\vskip -0.25 cm
\includegraphics[width=6cm,angle=270]{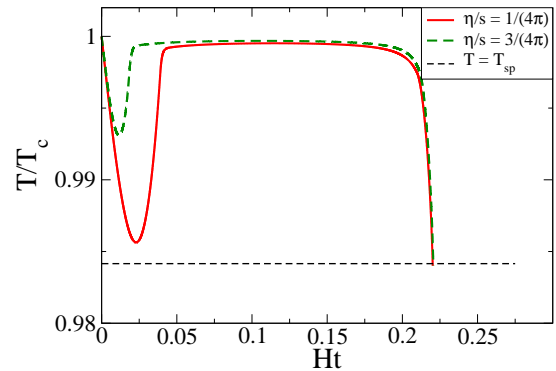}
\caption{Temperature, in units of $T_{c}$, as a function of time, in units of $1/H$ 
for different values of viscosity with $H^{-1}=600$ fm/c.}
\end{center}
\end{figure}

In Fig. 9 we can see that the EoS provided by the bag model, which leads to a latent heat 
larger than the one produced by the mixed EoS we adopt, implies a different transition 
dynamics. In fact, the larger latent heat of the bag EoS leads to a faster reheating of the 
plasma, since each nucleated bubble releases much more latent heat. Thus, a faster reheating 
means that the system will spend more time close to the critical temperature and the average 
nucleation rate will be much smaller for higher values of latent heat. Therefore, as can be 
seen in the figure, the nucleation process will take much longer to be completed.

\begin{figure}[!hbt]
\label{fig:reaq600_1_latt_bag}
\begin{center}
\vskip -0.25 cm
\includegraphics[width=6cm,angle=270]{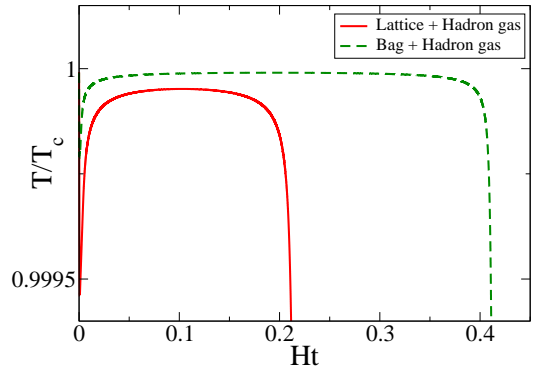}
\caption{Temperature, in units of $T_{c}$, as a function of time, in units of $1/H$, 
for the mixed and the bag model equations of state with $H^{-1}=600$ fm/c.}
\end{center}
\end{figure}

We can now focus on the hadronic fraction of the plasma, i.e., the fraction which has 
been hadronized via bubble nucleation up to time $t$, $f(t)$. As can be easily seen from 
Eq. (\ref{eq:frac_t}), $f(t)$ depends very strongly on $\tilde\delta\equiv 1-T/T_c$, another 
measure of the supercooling. In fact, for a small $\tilde\delta$, when the thin-wall approximation 
is valid, 
\begin{equation}
 \Gamma(T) = \frac{{\cal P}_0}{2\pi}\exp\left[{-\Delta F/T}\right] \approx \frac{{\cal P}_0}{2\pi} 
                 \exp{\left[-\frac{16\pi\sigma^3}{3\ell^2 T_c\tilde\delta^2}\right]} \;, 
\end{equation}
which depends very strongly on $\tilde\delta$. This implies that $f(t)$ will also grow 
exponentially as the temperature decreases, but before reheating effects are manifest. 
Indeed, after reheating the fluid temperature is near $T_c$ and it is unlikely that new 
bubbles nucleate. This means that after reheating $f(t)$ grows due to 
the expansion of the existing bubbles, i.e. in a much milder fashion than during the 
supercooling phase, as can be seen in Fig. 10. We also notice that, for a slow expansion, 
the first stage (nucleation of new bubbles) takes place very early, and almost all of the 
dynamics of phase conversion is due to the expansion of the bubbles. 

\begin{figure}[!hbt]
\label{fig:fraction_t}
\begin{center}
\vskip -0.25 cm
\includegraphics[width=6cm,angle=270]{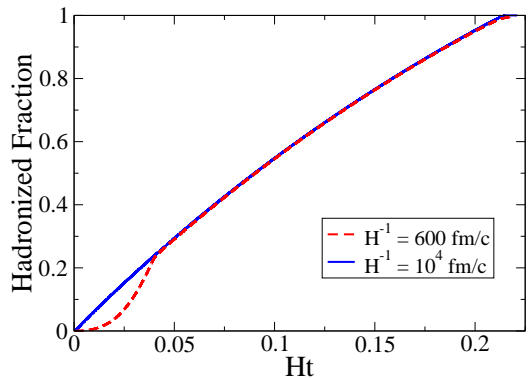}
\caption{Hadronized fraction of space as a function of time for $H^{-1}=600$ fm/c 
(critical expansion) and for $H^{-1}=10^4$ fm/c (slow expansion).}
\end{center}
\end{figure}

By examining both $T(t)$ and $f(t)$ on the same plot (Fig. 11), one can see quite clearly the 
relation between the reheating and the growth of $f(t)$ on one hand and, on the other hand, that 
after the reheating the temperature drops only after the phase transition is completed, i.e., 
after the system reaches $f=1$.

\begin{figure}[!hbt]
\label{fig:comparT_f}
\begin{center}
\vskip -0.25 cm
\includegraphics[width=6cm,angle=270]{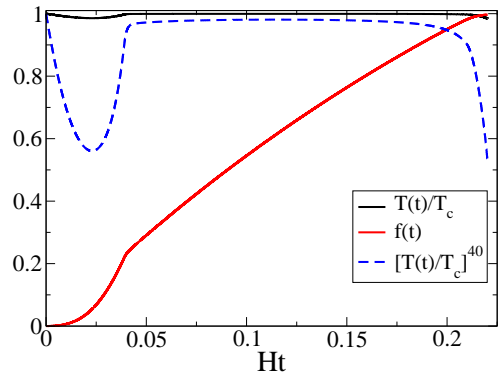}
\caption{$T(t)$ and $f(t)$ for $H^{-1}=600$ fm/c. Full lines 
are the actual $T(t)$ and $f(t)$, and the dashed line is a {\it zoomed} $T(t)$ for comparison 
with $f(t)$.}
\end{center}
\end{figure}

If the plasma is quenched directly into the spinodal region and its hadronization begins, there 
will also be a release of latent heat, just like in nucleation, and the system may also reheat. A 
curious possibility is that of a double process phase conversion, in which the reheating 
takes the plasma back to some metastable temperature, in which nucleation dominates. As 
a result, part of the plasma is converted through spinodal decomposition and part by 
(inhomogeneous) nucleation. It would also be interesting to explore experimental consequences 
of such a hybrid  transition process.

\section{Conclusions and outlook}

Using two different equations of state, a realistic matching of hadron gas of resonances and 
Lattice QCD for $N_f=2+1$ on one side, and the bag model (matched onto a hadron gas in 
the low temperature sector) on the other, we calculated both static and dynamical features of 
homogeneous bubble nucleation in a weakly first-order quark-hadron transition scenario, which 
is physically appealing if one takes into account results from the lattice and from experiments, 
as well as the nonequilibrium nature of the phase conversion process, especially in the case 
of high-energy heavy ion collisions. 

After setting up an effective potential, we obtained numerically bubble profiles, critical 
radii, the surface tension and the free energy as functions of the temperature. We also 
compared our numerical results to those derived in the thin-wall approximation. We showed 
how this approximation (valid only near the critical temperature) can be used to 
{\it overestimate} the nucleation rate for lower temperatures. This indicates that an 
adequate approach to the dynamics of nucleation away from $T_c$ should take into account 
exact bubble profiles $\phi(r)$, which in general have to be calculated numerically. With 
these {\it static} quantities we were able to feed a model for the dynamics of bubble nucleation 
in a homogeneously and isotropically expanding plasma, which is often adopted in studies of 
the early universe and heavy-ion collisions. Within this model, the only essential difference 
between these two physical settings is the value of the inverse Hubble constant $H^{-1}$, which 
is about $10^{19}$ fm/c in the early Universe during the quark-hadron transition era, and about 
$10$ fm/c in the case of heavy-ion experiments. We computed the temperature and the fraction of 
hadronized plasma as functions of time for different expansion rates, $H$, in order to make a 
quantitative estimate of the importance of nucleation in (homogeneous and isotropic) expanding 
systems. 

In the scenario of a weakly first-order transition investigated in this paper, it is clear that 
nucleation remains as the dominant mechanism of phase conversion in the early universe, as expected 
from the standard cosmological picture. Nevertheless, previous estimates for relevant time scales 
using the bag model equation of state, which yields a much stronger first-order transition, may differ 
by a factor of two when compared to results from a more realistic equation of state. For high-energy 
heavy ion collisions, where the plasma expands very quickly, the main mechanism for phase conversion 
must be greatly dominated by spinodal decomposition, which possibly has some effect on particle 
correlations and fluctuation \cite{polyakov,paech,Aguiar:2003pp,Sasaki:2007db,Mishustin:1998eq,
Heiselberg:2000ti,Bower:2001fq,Pruneau:2002yf,Chomaz:2003dz,Misshustin:2007jx}. This issue 
will be addressed in a future publication.

\acknowledgements
The authors thank T. Kodama for fruitful discussions. E.S.F. is grateful to T. Mendes for 
many useful comments and discussions. 
This work was partially supported by CAPES, CNPq, FAPERJ, FAPESP, and FUJB/UFRJ. 


\end{document}